\documentclass[submission,copyright,creativecommons,noderivs,noncommercial]{eptcs}
\providecommand{\event}{WWV 2011}

\usepackage{times}
\usepackage{latexsym }
\usepackage{amssymb}
\usepackage{amsmath}
\usepackage{array}
\usepackage{graphicx}
\usepackage{algorithm}
\usepackage{algorithmic}
\usepackage{listings}
\usepackage{multicol}
\usepackage{pifont}
\usepackage{subfigure}
\newtheorem{example}{Example}
\newtheorem{definition}{Definition}
\newcommand{\hspt}{\hspace*{0.25in}}
\begin{document}

%\author{Naseem Ibrahim}
\title{Specification and Verification of Context-dependent Services}
%\institute{Concordia University, Montreal, Canada\\n\_ibrah@encs.concordia.ca}
\author{Naseem Ibrahim, Vangalur Alagar, and Mubarak Mohammad
\institute{Department of Computer Science \& Software Engineering\\
Concordia University, Montreal, Canada}
\email{\{n\_ibrah,alagar,ms\_moham\}@cse.concordia.ca}}
\def\titlerunning{Spec. and Verif. of Context-dependent Services}
\def\authorrunning{N. Ibrahim, V. Alagar, M. Mohammad}

\maketitle

%\tableofcontents
\begin{abstract}
Current approaches for the discovery, specification, and provision of services ignore the relationship between the service contract and the conditions in which the service can guarantee its contract. Moreover, they do not use formal methods for specifying services, contracts, and compositions. Without a formal basis it is not possible to justify through formal verification the correctness conditions for service compositions and the satisfaction of contractual obligations in service provisions. We remedy this situation in this paper. We present a formal definition of services with context-dependent contracts. We define a composition theory of services with context-dependent contracts taking into consideration functional, nonfunctional, legal and contextual information. Finally, we present a formal verification approach that transforms the formal specification of service composition into extended timed automata that can be verified using the model checking tool UPPAAL.
\end{abstract}
%\tableofcontents
%-------------------------------------------------
\section{Introduction}
In~\cite{NI11} and~\cite{SCC211}, we introduced a formal framework, called \emph{FrSeC}, that supports the specification, publication, discovery, selection, composition and verification of services with context-dependent contracts. The work reported in this paper is founded on this framework. We provide a formal specification of services with context-dependent contracts and their compositions. The composition theory of services takes into consideration the functional, nonfunctional, legal, and contextual aspects of services. We also present a formal verification approach that transforms the formal specification of service composition into UPPAAL~\cite{tut04} timed automata in order to verify service properties using model checking.

\emph{Service-oriented Architecture (SOA)} is an emerging view of the future of distributed computing and enterprise application development~\cite{1296147}. However, current approaches for the specification, publication, discovery, selection, and provision of services fall short in important respects. First, the relationship between the service contract and the conditions in which the service can guarantee its contract has been ignored, however these are necessary in order to associate the context of the service provider and the context of the service requester. Second, contextual information~\cite{593572} is not well represented and not rigorously applied in service discovery and service provision. Third, current composition approaches compose only service functionality and ignore nonfunctional requirements. Thus, service contracts, and context information that are part of services are left out of the composition, and verification. Fourth and the last, the published approaches do not use formal methods for the specification of services, contracts, contextual representation and application, and service composition. Without a formal basis it is not possible to justify through formal verification the correctness conditions for service compositions and the satisfaction of contractual obligations in service provisions. The work reported in this paper eliminates these shortcomings.

The basic building unit for SOA-based applications is \emph{service}. It is normally understood that service is an autonomous and platform-independent software program, having its own distinct functionality and a set of capabilities related to this functionality. These capabilities are usually invoked by external consumer programs and are usually expressed via a published \emph{service contract}. A service contract establishes the terms of engagement with the service, provides technical constraints and requirements, and provides any semantic information the service owner wishes to make public~\cite{1296147}. We reckon that a Service Provider will package service functionality with non-functional attributes, service contract, and context. So, we decided to deal with \emph{ConfiguredServices}, which are formalized in Section~\ref{sec:cs}. A Service Provider may choose to compose \emph{ConfiguredServices}. The composition mechanism itself may be driven by the business model of the Service Provider. Keeping this point of view, we discuss in Section~\ref{sec:comp} a formal composition theory of services (\emph{ConfiguredServices}). Section~\ref{sec:verify} presents an approach to formally verify service properties in service compositions. In Section~\ref{sec:related}, we briefly, yet critically, compare our work with related work. Finally, Section~\ref{sec:conc} concludes the paper with a summary of ongoing work.

\section{\emph{ConfiguredService} Definition}\label{sec:cs}
\begin{figure}
	\centering
		\includegraphics[width=0.7\textwidth]
		{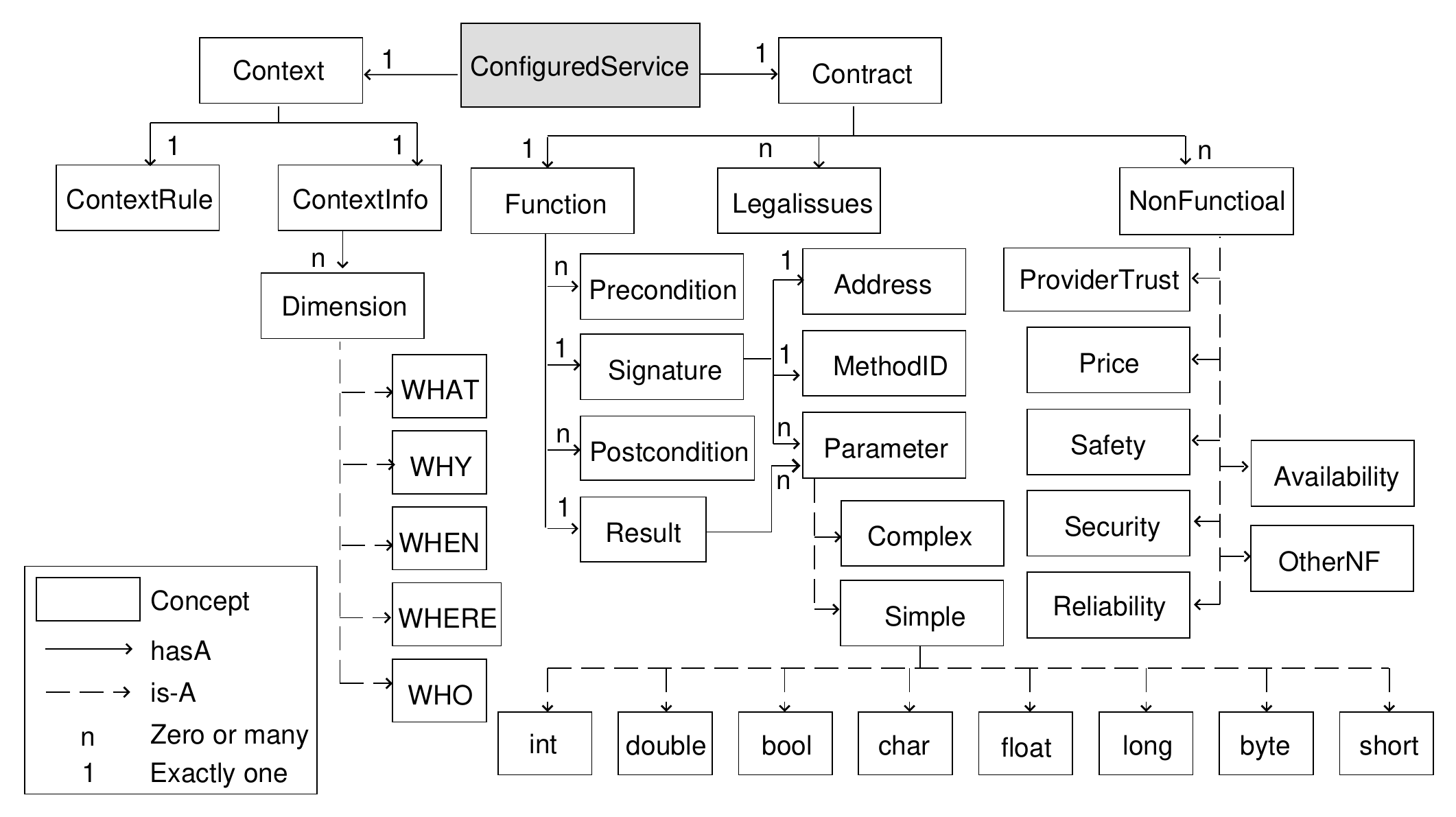}
	\caption{\emph{ConfiguredService}}\label{Fig:cs}
\end{figure}

Services are defined by service providers in a contract first approach. That is, the contract is defined before the implementation of service~\cite{1296147}. The service provider determines all the possible contracts that this service should satisfy. Then, the service provider defines the \emph{ConfiguredServices} that represents those contracts. After that, the service provider develops the \emph{ImplementedService} that implements the different \emph{ConfiguredServices} that provide the same functionality with different contracts and contexts. A \emph{ConfiguredService} is to be published in Service Registry and made available for discovery and selection. A \emph{ConfiguredService} is a package in which service functionality, service contract, and service provision context are bundled together. The Service Provider publishes the \emph{ConfiguredService} elements. The published elements should be sufficient for the discovery and selection of this service. The essential elements that make this happen are \emph{contract} and \emph{context}, as shown in Figure~\ref{Fig:cs}. The contract will include \emph{function}, \emph{nonfunctional properties} and \emph{legal issues}. Trustworthiness features are included in the nonfunctional part of the contract and legal issues include business rules and other trade laws within the context. 
\begin{itemize}
\item \textbf{Function}: A \emph{ConfiguredService} provides a single function. The function definition will include the function \emph{signature}, \emph{result}, \emph{preconditions} and \emph{postconditions}. The \emph{signature} part defines the function \emph{identifier}, the invocation \emph{address}, and the \emph{parameters} of the function. Each parameter has an \emph{identifier} and a \emph{type}. The \emph{result} part defines the returned data of the service function. The \emph{preconditions} define the conditions that should be true before the function invocation. The \emph{postconditions} define the conditions that are guaranteed to be true after the function invocation. 

\item \textbf{Nonfunctional properties}: A \emph{ConfiguredService} definition includes nonfunctional properties that it can guarantee. These properties are to be chosen carefully so that they are verifiable, and encompass both quality and quantity aspects of service. \emph{Trustworthiness} and {\em Price} are examples. Trust itself is further divided into \emph{ConfiguredService trust} and \emph{provider trust}. These are explained in detail in the next section. \emph{ConfiguredService trust} defines the trustworthiness properties that are related to service provision. It includes the features \emph{safety}, \emph{security}, \emph{availability}, and \emph{reliability}~\cite{Mohammad201177}. Safety defines the critical conditions that are guaranteed to be true by Service Providers, such as timing conditions. Security is a composite of data integrity and confidentiality. Availability can be defined as the extent of readiness for providing correct services. Availability is specified as the maximum accepted time of repair until the service returns back to operate correctly. Reliability is the quality of continuing to provide correct services despite a failure. It is defined as the accepted mean time between failures. \emph{Provider trust} defines the trustworthiness properties that are related to the Service Provider. It may include recommendations from other clients, and lowest prices guarantees. There is no agreed upon definition for Provider trust. The main issue here is the inclusion of verifiable information that makes a seller trusted. 

\item \textbf{Legal Issues}: One of the essential elements of the \emph{ConfiguredService} contract is the set of legal rules that constrain the contract. Business rules, such as \emph{refund conditions}, \emph{interest and administrative charges}, and \emph{payment rules}, form one part of legal issues. Another part is the set of trade laws enforced in the context of service request and delivery. Examples of the later kind are \emph{service requesters rights}, \emph{privacy laws}, and \emph{censor rules}. In the literature~\cite{sul07}, no distinction was made between legal rules and nonfunctional requirements. We reckon that a clear distinction should be made between legal rules and nonfunctional properties. In many situations, if a nonfunctional property is `a soft' requirement it may be ignored, however ignoring a legal rule is equivalent to `legal violation', which might land in legal disputes and even lead to loss of entire business. In essence, not enforcing a legal rule prevents the execution of a contract. 
\end{itemize}

The context part of the \emph{ConfiguredService} will include the main parts \emph{ConfiguredService context} and \emph{context rules}. The \emph{ConfiguredService context} defines the contextual information of the \emph{ConfiguredService}. Context is formally defined in~\cite{wan06} using {\em dimensions} and {\em tags} along the dimensions. We illustrate context specification using the three dimensions \emph{WHERE}, \emph{WHEN} and \emph{WHO}. The dimension {\em WHERE} is associated with a location, which may be one or more of $\{$\emph{Point}, \emph{Region}, \emph{Address}, \emph{Route}, \emph{URI}, \emph{IP}$\}$. The dimension {\em WHEN} is associated with temporal information, such as time and date. The dimension {\em WHO} is associated with subject identities, such as the names of Service Providers and Service Requesters. We can also use {\em WHO} dimension to associate information from job roles. The \emph{context rules} define the contextual information related to the Service Requester that should be true for the Service Provider to guarantee the contract associated with the \emph{ConfiguredService}. Rules are defined as constraints in a subset of {\em Timed Computation Tree Logic} (TCTL), the logic used in UPPAAL. In practice, constraints can be expressed as simple logical expressions within the first order predicate logic (FOPL), a subset of TCTL.

\begin{example}\label{ex:zero}
This example introduces a simplified case study~\cite{terbeek08}, restricted to emergency \emph{road assistance service scenarios for automobiles}. A typical scenario is the breakdown of a car on a highway, which requests for road-side assistance. The car sends information to the nearest road assistance center, which in turn will use the information received to identify the repair shop, tow truck and car rental companies in that zone. In this example we identify three \emph{ConfiguredServices}, whose detailed definitions are shown in Figure~\ref{fig:example4-cs}.
\begin{figure}[!h]
	\centering
		\includegraphics[width=1.0\textwidth]
		{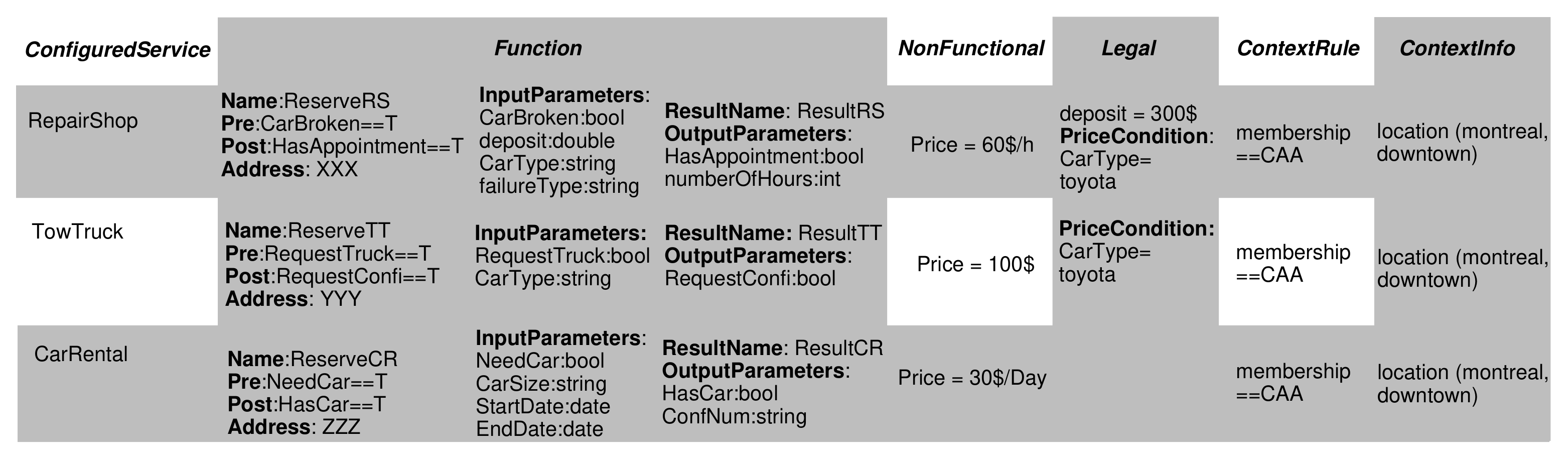}
	\caption{Roadside Emergency Services: \emph{ConfiguredServices} Description}\label{fig:example4-cs}
\end{figure}
\end{example}

\subsection{Formal Notation}
We use a model-based approach to formally specify \emph{ConfiguredServices}. The models are built from set theory and logic. The model is built incrementally, according to the template in Figure~\ref{Fig:cs}.

\noindent {\bf Definition of Constraints}: A constraint is a logical expression, defined over data parameters and attributes. Any well-formed formula built by using standard logical operators, quantifiers, and temporal operators allowed in TCTL~\cite{tut04} is a valid constraint. If $\mathbb{C}$ denotes the set of all such logical expressions, $X \in \mathbb{C}$ is a constraint. The following notation is used in our definition:
\begin{itemize}
\item $\mathbb{T}$ denotes the set of all data types, including abstract data types.
\item $Dt\in \mathbb{T}$ means $Dt$ is a datatype.
\item $v:Dt$ denotes that $v$ is either constant or variable of type $Dt$.
\item $X_v$ is a constraint on $v$. If $v$ is a constant then $X_v$ is true.
\item $V_{q}$ denotes the set of values of data type $q$.
\item $x::\Delta$ denotes a logical expression $x \in \mathbb{C}$ defined over the set of parameters $\Delta$.
\end{itemize}

\noindent {\bf Definition of Parameters}: A parameter is a 3-tuple, defining a data type, a variable of that type, and a constraint on the values assumed by the variable. We denote the set of data parameters as $\Lambda = \{\lambda=(Dt,v,X_v)|Dt\in\mathbb{T},v:Dt,X_v\in \mathbb{C}\}$. 

\noindent {\bf Definition of Attributes}: An attribute has a name and type, and is used to define some semantic information associated with the name. As an example, each \emph{ConfiguredService} can be given a version number, which is defined as an attribute. The set of attributes is $\alpha = \{(Dt,v_{\alpha})|Dt\in \mathbb{T}, v_{\alpha}:Dt\}$.

\noindent {\bf Definition of Context:} A context is formalized as a 2-tuple $\beta$ = $\langle r, c\rangle$, where $r \in \mathbb{C}$, built over the contextual information $c$. Context information is formalized using the notation in~\cite{wan06}: Let $\tau$ : $DIM$ $\rightarrow$ $I$, where $DIM$ = \{$X_1$, $X_2$,...,$X_n$\} is a finite set of dimensions and $I = \{a_1, a_2,..., a_n\}$ is a set of types. The function $\tau$ associates a dimension to a type. Let $\tau$($X_i$) = $a_i$, $a_i$ $\in$ $I$. We write $c$ as an aggregation of ordered pairs ($X_j$, $v_j$), where $X_j$ $\in$ $DIM$, and $v_j$ $\in$ $\tau$($X_j$).

\noindent {\bf Definition of Contract:} A contract is a 3-tuple $\sigma$ = $\langle f, \kappa, l \rangle$, where the service function $f$, the set of nonfunctional properties $\kappa$ and the set $l$ of legal issues that bind the service contract are defined below.

\begin{itemize}
\item {\em Service Function:}
A service function is a 4-tuple $f = \langle g, i, pr, po\rangle$, where $g$ is the function signature, $i$ is the function result, $pr$ is the precondition, and $po$ is the postcondition. A signature is a 3-tuple $g = \langle n, d, u\rangle$, where $n=(x|x:string)$ is the function identification name, $d=\{x|x \in \Lambda\}$ is the set of function parameters and $u=(x|x:string)$ is the function address, the physical address on a network that can be used to call a function. For example, it can be an IP address. The result is defined as $i = \langle m, q\rangle$, where $m=(x|x:string)$ is the result identification name and $q=\{x|x \in \Lambda\}$ is the set of parameters resulting from executing the \emph{ConfiguredService}. The precondition $pr$ and postcondition $po$ are data constraints. That is, $pr = \{y | y::z , z \subseteq \Lambda\}$ and $po = \{y| y::z, z \subseteq \Lambda\}$. 
\item {\em Nonfunctional Property:}
A nonfunctional property of a \emph{ConfiguredService} is a composite property, written as a 6-tuple $\kappa = \langle \rho, \epsilon, \psi, \eta, p, tr \rangle$, where $\rho$ is the safety guarantee, $\epsilon$ is the security guarantee, $\eta$ is the availability guarantee, $\psi$ is the reliability guarantee, $p$ is the service cost and $tr$ is a measure of the provider trust. The safety guarantee includes time guarantee $\rho_t$ and data guarantee $\rho_d$. We assume that $time$ is a generic type. The time guarantee is defined as $\rho_t = (x | x:time)$, the time the service takes to provide its function. The data guarantee refers to the accuracy of data, and is defined as $\rho_d = \{x| x::z, z \subseteq \Lambda\}$. Let $H$ denote the set of security protocols that the Service Provider has followed to guarantee confidentiality and integrity constraints. Then the set $\epsilon = \{x | x \in H\}$ defines the extent of security binding the service. The reliability guarantee refers to the guaranteed maximum time between failures, and is defined as $\psi = (x | x:time)$. The availability guarantee refers to the guaranteed maximum time for repairs, and is defined as $\eta = (x | x:time)$. The price is defined as a 3-tuple $p = \langle a, cu ,un\rangle$, where $a = (x|x:\mathbb{N})$ is the price amount defined as a natural number, $cu=(y|y:cType)$ is currency tied to a currency type $cType$, and $un=(z|z:uType)$ is the unit for which pricing is valid. As an example, $p = (100, \$, hour)$ denotes the pricing of $100\$$/hour. Provider Trust is defined as a 3-tuple $tr=\langle ce, pg, re\rangle $, where $ce$ is recommendations from other clients, $pg$ is lowest prices guarantees and $re$ is recommendations from independent organizations. Lowest price guarantee is represented by a flag $pg=(a| a:Boolean)$. It is a Boolean that is true when a \emph{ConfiguredService} can guarantee its price to be lower than the price of any other \emph{ConfiguredService} providing the same functionality. Client recommendations and recommendations from independent organizations can be defined as sets of ordered pairs. In $ce = \{(a,b)|a:CLIENT, b \in \{Low, Below Average, Average, Above Average, High\}\}$, a pair $(a,b)$ represents a client $a$ whose recommendation is $b$. Likewise, in $re = \{(a,b)|a:ORGANIZATION, b \in \{Low, Below Average, Average, Above Average, High\}\}$, a pair $(a,b)$ represents an organization $a$ whose recommendation is $b$.
\item {\em Legal Issues:}
A legal issue is a rule, expressed as a logical expression in $\mathbb{C}$. A rule may imply another, however no two rules can conflict. We write $l = \{y| y \in \mathbb{C}\}$ to represent the set of legal rules.
\end{itemize}
\noindent Putting these definitions together we arrive at a formal definition for \emph{ConfiguredService}.
\begin{definition}
A \emph{ConfiguredService} is a 4-tuple $s$ = $\langle \Lambda, \alpha, \beta, \sigma \rangle$, where $\Lambda$ is a set of parameters, $\alpha$ is a set of attributes, $\beta$ is a context, and $\sigma$ is a contract.
\end{definition}
\noindent We remark that not all components of $\kappa$ may be relevant for a service, as shown in many later examples. In general, the trust domain, in which $ce$ and $pg$ are defined, must be a {\em complete lattice}~\cite{Wal08}. This property is essential in order to compare trust values of groups and compute minimum (maximum) among trust values. For the sake of simplicity, we assume in further discussion that trust values assumed by $ce$ and $re$ are whole numbers in the range $1 \ldots 5$, where $1$ denotes $Low$ and $5$ denotes $High$. This assumption will enable us to calculate simple averages, maximum, and minimum of a set of trust values. Example~\ref{ex:first} illustrates the application of the above formal notation to the \emph{ConfiguredService} defined in Example~\ref{ex:zero}.
\begin{example}\label{ex:first}
Let $rs$ denote the ConfiguredService for providing a Repair Shop who provides the services described in Figure~\ref{fig:example4-cs}. The formal notation of \emph{ConfiguredService} $rs$ is $s_{rs}$ = $\langle \Lambda_{rs}, \alpha_{rs}, \beta_{rs}, \sigma_{rs},  \rangle$, where the tuple components are explained below.
\begin{itemize}
\item parameters: $\Lambda_{rs}=\{(CarBroken,bool),(deposit,double),(CarType,string),(failureType,string),$ $(HasAppointment,bool),(numberOfHours,int)\}$.
\item attributes: $\alpha_{rs}=\Phi$.
\item context: $\beta_{rs}$ = $\langle r_{rs}, c_{rs}\rangle$, where $r_{rs} = \{(membership==caa)\}$ is the context rule and $c_{rs}$ $= \{(Location, (Montreal, Canada))\}$ is the contextual information of the emergency road service provider 
\item contract: $\sigma_{rs}$ = $\langle f_{rs}, \kappa_{rs}, l_{rs} \rangle$, where the elements of the 3-tuple are defined below:\\
\hspt 1. contract functionality specification: $f_{rs} = \langle g_{rs}, i_{rs}, pr_{rs}, po_{rs}\rangle$\\
\hspt\hspt 1.1 function signature: $g_{rs} = \langle n_{rs}, d_{rs}, u_{rs} \rangle$, where \\
\hspt\hspt\hspt $n_{rs} = (ReserveRS)$ is the name, $d_{rs} = \{(CarBroken,bool),$ $(deposit,$ $double),$\\
\hspt\hspt\hspt $(CarType,string),(failureType,string)\}$ are input data parameters, and $u_{rs} =(XXX)$\\
\hspt\hspt\hspt  is the address\\ 
\hspt\hspt 1.2 function result: $i_{rs} = \langle m_{rs}, q_{rs}\rangle$ , where \\
\hspt\hspt\hspt $m_{rs} = (ResultRS)$ is the name and the set of output data parameters is\\
\hspt\hspt\hspt $q_{rs} =\{(HasAppointment,bool),(number$ $OfHours,int)\}$\\
\hspt\hspt 1.3 function precondition: $pr_{rs} = \{(CarBroken==true)\}$ \\
\hspt\hspt 1.4 function postcondition $po_{rs}=\{(HasAppointm$ $ent==true)\}$\\
\hspt 2. contract nonfunctional property specification: $\kappa_{rs} = \langle p_{rs} \rangle$, $p_{rs} = \langle a_{rs}, cu_{rs} ,un_{rs}\rangle$, where 
\hspt $a_{rs} = (60)$ is the cost, $cu_{rs} = (dollar)$ is the currency, and $un_{rs} = (hour)$ is the pricing unit\\
\hspt 3. contract legal issue specification: $l_{rs} = \{(deposit=300),(CarType==toyota)\}$, where
\hspt the deposit amount is $300$ and the car type is $toyota$.
\end{itemize}
\end{example}

\section{Service Composition}\label{sec:comp}
Although service composition has been considered before by some researchers~\cite{1296147,StarWSCop} no specific method has been put forth. In \emph{FrSec} a service composition may be attempted either at design-time or at execution-time. The former, called \emph{static} composition, is driven by Service Provider's business goals. The later, called \emph{dynamic} service composition, is driven by user's demands at service provision contexts. In this paper, we focus only on static service composition. We present a few composition constructs, give their semantics and suggest a verifiable composition theory. 

\subsection{Composition Constructs}
Defining a composite service includes defining the execution logic of the participant services. This section, inspired by~\cite{WMA09}, defines the composition constructs and informally motivates their execution logics. 
\begin{figure}[!h]
\begin{center}
	\subfigure[]{\label{fig:seq}\includegraphics[width=0.08\textwidth]{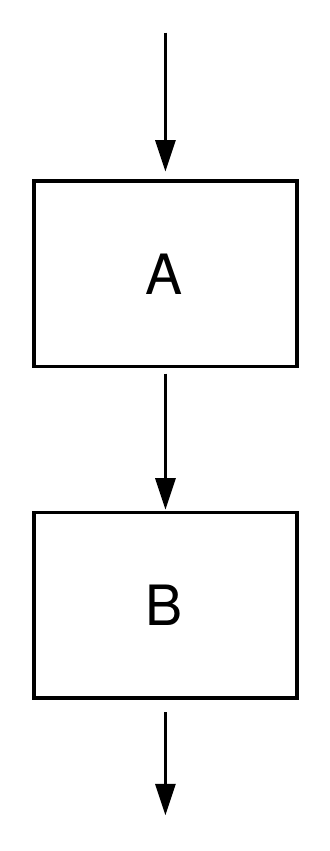}}
	\subfigure[]{\label{fig:para}\includegraphics[width=0.15\textwidth]{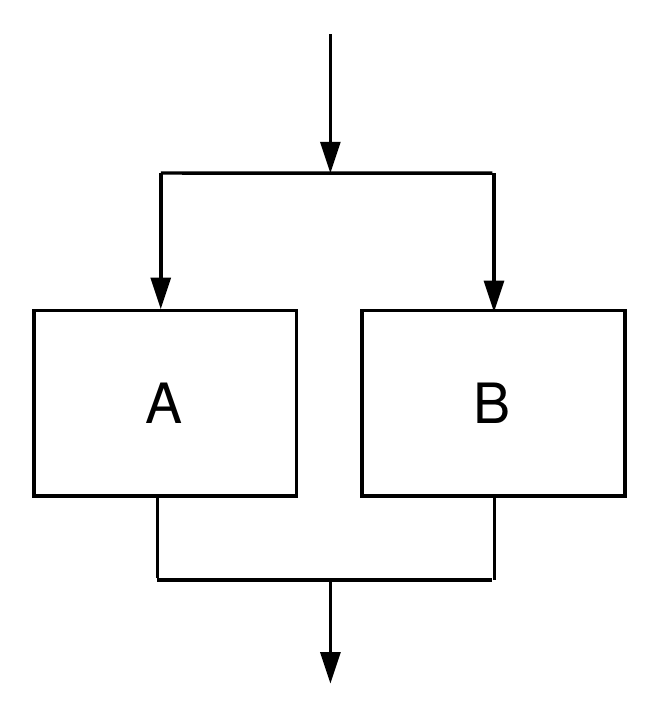}}
	\subfigure[]{\label{fig:priority}\includegraphics[width=0.1\textwidth]{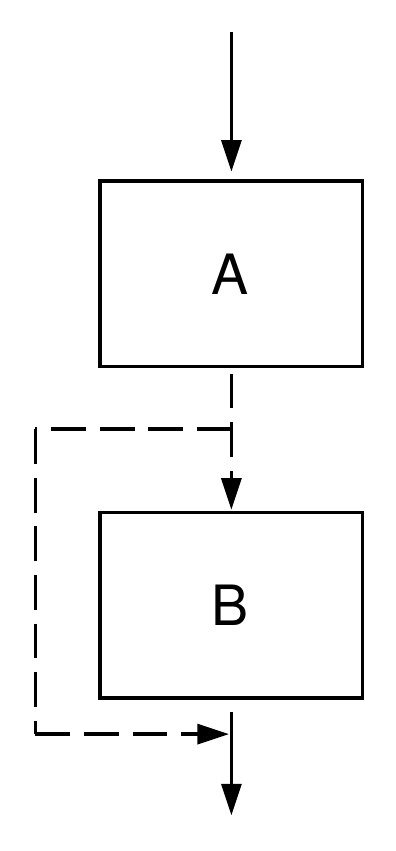}}
	\subfigure[]{\label{fig:noorder}\includegraphics[width=0.15\textwidth]{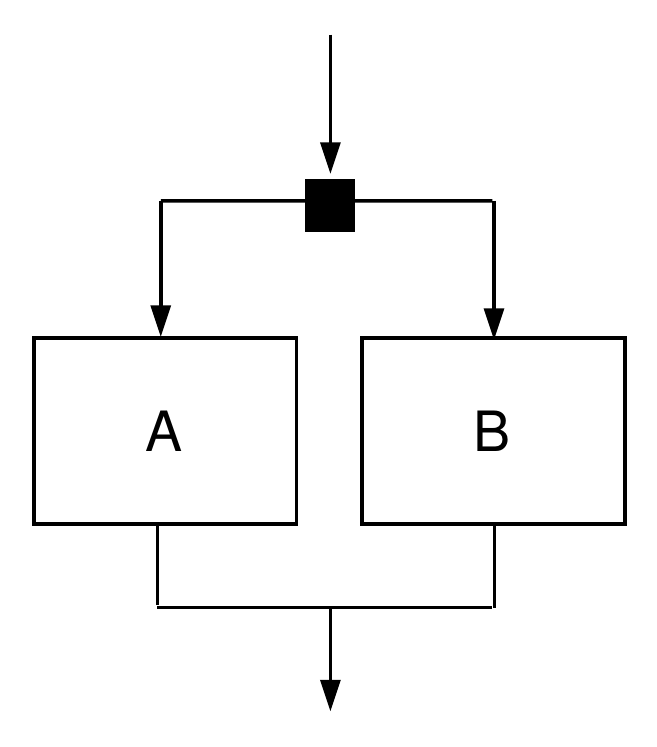}}
  \subfigure[]{\label{fig:nondet}\includegraphics[width=0.15\textwidth]{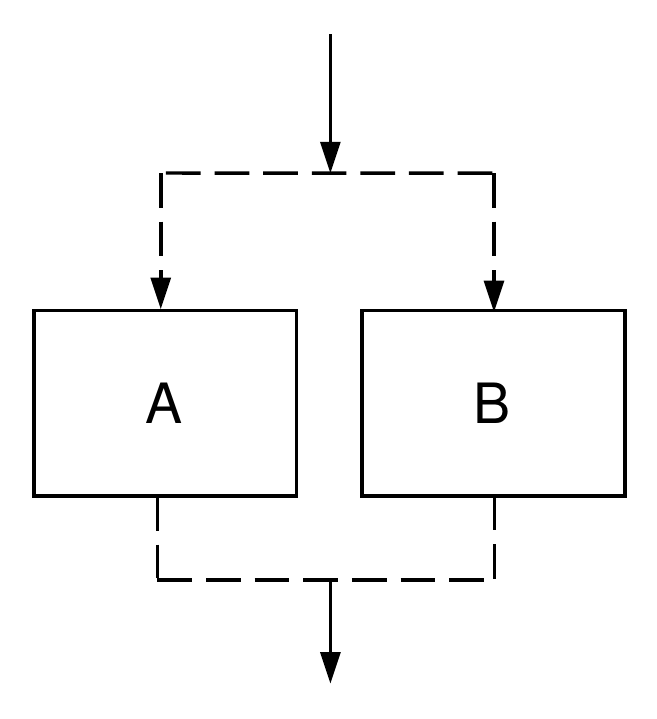}}
	\subfigure[]{\label{fig:if}\includegraphics[width=0.15\textwidth]{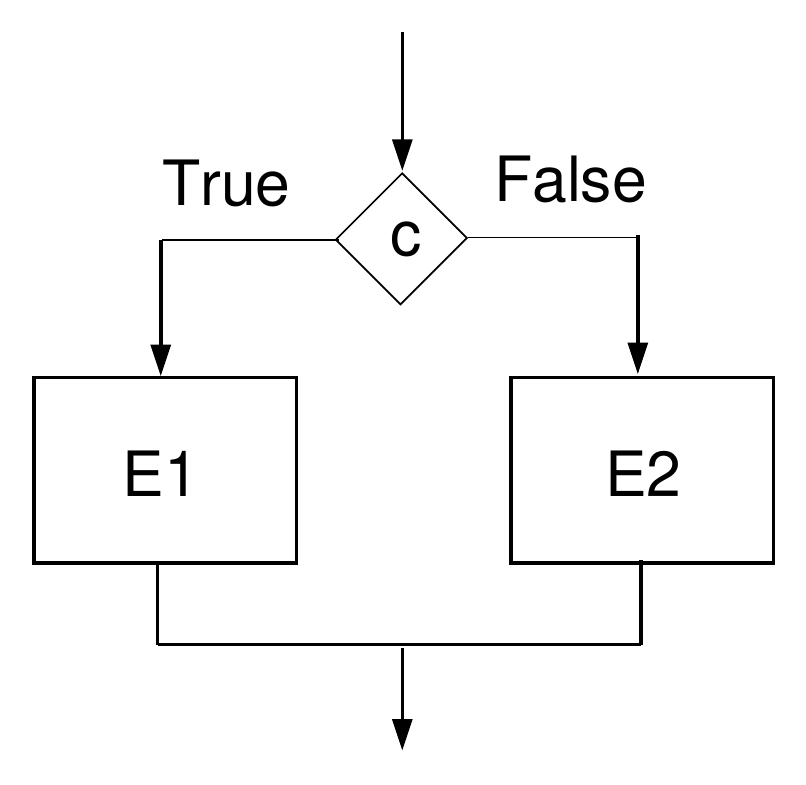}}
	\subfigure[]{\label{fig:loop}\includegraphics[width=0.15\textwidth]{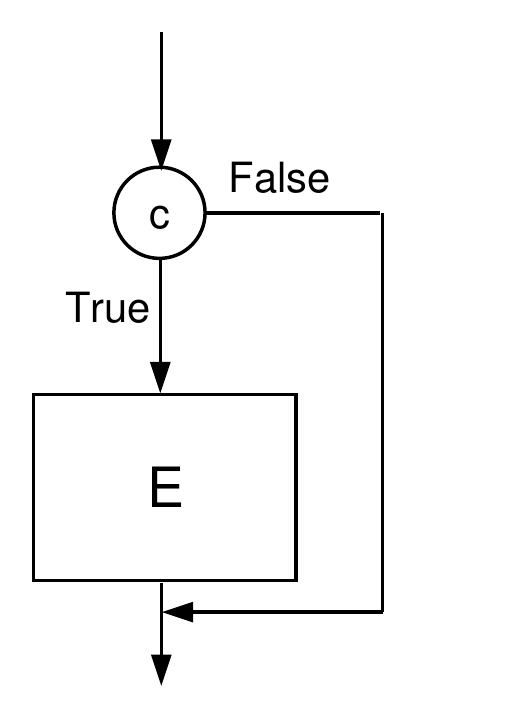}}
\end{center}
  \caption{a) Sequential, b) Parallel, c) Priority d) No order, e) Nondeterministic, f) Conditional, and g) Iteration Constructs}
  \label{fig:edge}
\end{figure}
\begin{itemize}
\item \textbf{Sequential composition construct $\gg$}: Given two \emph{ConfiguredServices} $A$ and $B$, the expression $A \gg B$ (Figure~\ref{fig:seq}) defines the sequential composition of $A$ and $B$. The execution logic of this composite service is that \emph{ConfiguredService} $A$ is executed first, and its output may be used in the execution of \emph{ConfiguredService} $B$, in addition to any input that $B$ may require. In general, the expression $A_1 \gg A_2 \ldots \gg A_k$ denotes the execution of \emph{ConfiguredService} $A_{i+1}$ with the result of execution of $A_i$ as an input, for $i= 1, \ldots, k-1$, in addition to other input that $A_{i+1}$ might need.

\item \textbf{Parallel composition construct $||$}: Given two \emph{ConfiguredServices} $A$ and $B$, the expression $A||B$ (Figure~\ref{fig:para}) defines the parallel composition of $A$ and $B$. The parallel composition $A || B$ models the simultaneous executions of \emph{ConfiguredServices} $A$ and $B$. In general, the evaluation of the expression $A_1 \parallel A_2 \parallel \ldots \parallel A_k$ will create $k$ service execution threads, one for each \emph{ConfiguredService}. The result of this composite service is the merging of their individual results in time order. That is, the execution of the composite service finishes only when all service executions terminate.

\item \textbf{Priority construct $\prec$}: Given two \emph{ConfiguredServices} $A$ and $B$, the expression $A \prec B$ (Figure~\ref{fig:priority}) defines that the service execution of $A$ should be attempted first, and if it succeeds, the service $B$ is to be discarded; otherwise, the execution of service $B$ should be attempted. In general, the expression requires that the service executions be attempted deterministically in the order specified until the first successful execution of service. The meaning of the expression $A_1 \prec \ldots \prec A_k$ is that the service that can be successfully executed is the result of the composition.

\item \textbf{Composition with no order $\Diamond$}: Given two \emph{ConfiguredServices} $A$ and $B$, the expression $A \Diamond B$ (Figure~\ref{fig:noorder}) defines that services $A$ and $B$ should be executed by the receiver, however the order of their executions is not important. The result of the composition is the set of results produced by the executions of the \emph{ConfiguredServices} $A$ and $B$. In general, the expression $A_1 \Diamond A_2 \Diamond \ldots \Diamond A_k$ defines the composition of services $A_i$, $i= 1,k$ when all of them may be executed in no specific order.

\item \textbf{Nondeterministic choice construct $\wr$}: Given two \emph{ConfiguredServices} $A$ and $B$, the expression $A \wr B$ (Figure~\ref{fig:nondet}) defines the composition in which one of the services is executed nondeterministically. In general, $A_1 \wr \ldots \wr A_k$ denotes the execution of a nondeterministically chosen service from the $k$ operands. If the service $A_i$ is the nondeterministic choice, the result from the evaluation of the service $A_i$ is the result of evaluating the composition $A_1 \wr \ldots \wr A_k$. In using this composition it is understood that any service $A_i$ can be chosen for the intended purpose.

\item \textbf{Conditional choice construct (if-else) $\triangleright$}: Given two service expressions $E_1$ and $E_2$, the composition expression $E_1\triangleright_c E_2$ (Figure~\ref{fig:if}) states that if condition $c$ evaluates to true then expression $E_1$ is to be chosen for execution, otherwise expression $E_2$ should be executed. 

\item \textbf{Iteration construct (while) $\circ$}: The composition $E_{\circ_c}$ (Figure~\ref{fig:loop}) states that the service expression $E$ should be executed as long as $c$ evaluates to true.
\end{itemize}

\noindent {\bf Construct Binding} All constructs have the same precedence, and hence a composite service expression is evaluated from left to right. To enforce a particular order of evaluations, parenthesis may be used.

\begin{figure}[h!]
	\centering
		\includegraphics[width=0.45\textwidth]
		{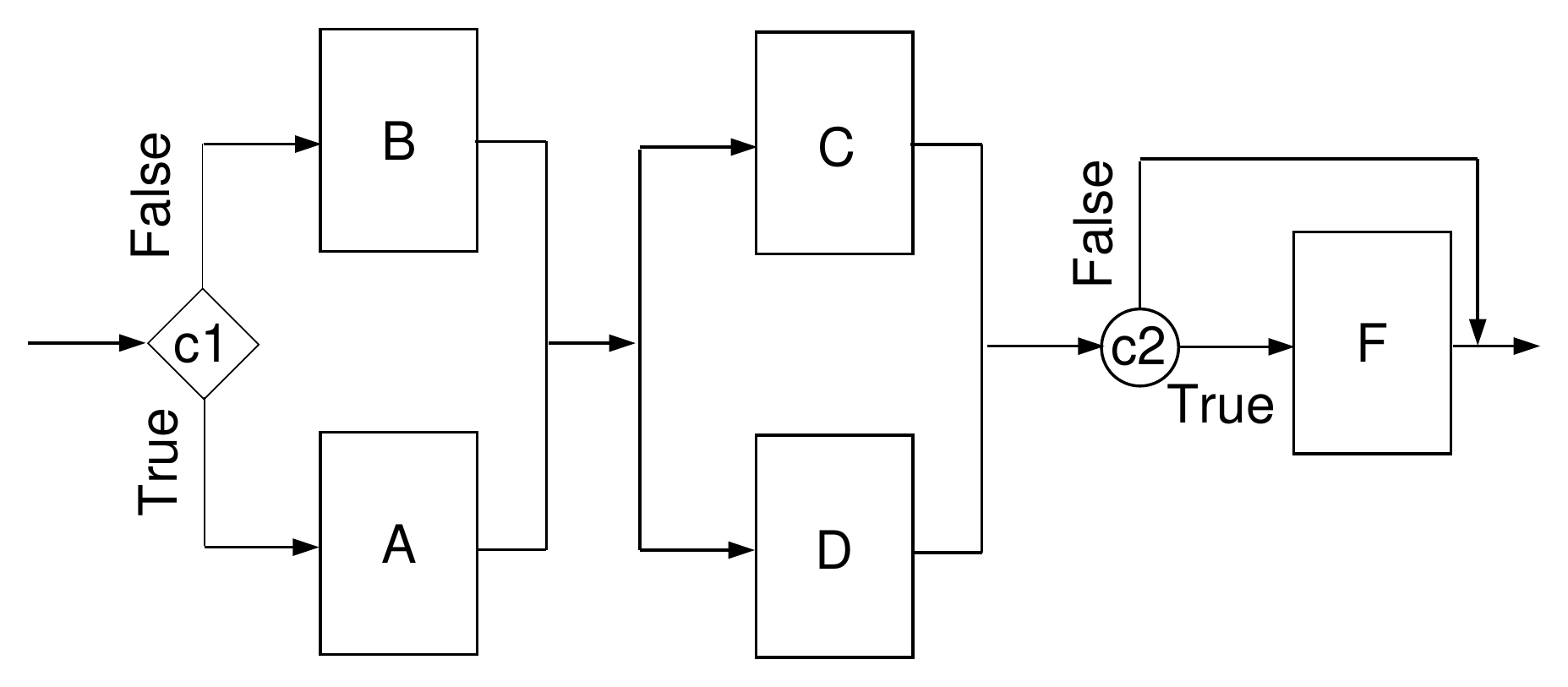}
	\caption{Execution logic of $(A\triangleright_{c1} B)\gg (C||D)\gg F_{\circ_{c2}}$}\label{fig:example1}
\end{figure}

\begin{example}\label{ex:second}
The execution logic of the composite service $(A\triangleright_{c1} B)\gg (C||D)\gg F_{\circ_{c2}}$, shown in Figure~\ref{fig:example1}, is obtained by putting together the execution logics from Figure~\ref{fig:edge}.
\end{example}

\subsection{Semantics of \emph{ConfiguredService} Compositions}
Every Service Provider has a business model. Motivated by the business rules and logic in the model, a Service Provider will determine the nature of composition for services. We want to emphasize that the {\em meaning} of a composition primarily rests on the chosen business goals and rules. Consequently, service compositions are very much {\em unlike} action compositions based purely on preconditions and postconditions. As an example, a Service Provider may form $A \gg B$ because service $B$ can be provided only after service $A$ has been provided. That is, service $B$ cannot be realized without first executing service $A$. This is analogous to `bootstrapping' before invoking any other system function in the domain of computing services. This implies that the precondition for invoking a system function includes the precondition for invoking `bootstrapping', however it might require more conditions to be met. Moreover, the postcondition of `bootstrapping' and the postcondition of the system function invoked after that are both observed. In some domains, it might happen that the precondition for invoking service $B$ is exactly the same as the postcondition of the first service $A$, and is not observable. Only the postcondition of $B$, after $B$ is completed, may be observable. Given such subtle scenarios, it is hard to give one `fixed' semantics for service compositions. Below we give the semantics for sequential composition. An account of the full semantics can be found in~\cite{NAM11}. The proposed semantics is appropriate for one kind of business logic, and our approach can be used to provide semantics for different business logics. By providing an approach to formal semantics for composition constructs we are motivating how a theory of composition can be developed.

Below we let
$A = \langle \Lambda_A, \alpha_A, \beta_A, \sigma_A \rangle$, and $B = \langle \Lambda_B, \alpha_B, \beta_B, \sigma_B \rangle$
denote two \emph{ConfiguredServices}, where $\beta_A = \langle r_A,c_A \rangle$, $\beta_B = \langle r_B,c_B \rangle$, 
$\sigma_A = \langle f_A, \kappa_A, l_A \rangle $, $\sigma_B = \langle f_B, \kappa_B, l_B \rangle $, $f_A = \langle g_A, i_A, pr_A, po_A \rangle$, $f_B = \langle g_B, i_B, pr_B,$ $po_B \rangle$, $g_A = \langle n_A, d_A, u_A \rangle$,  $g_B = \langle n_B, d_B, u_B \rangle$, $i_A = \langle m_A, q_A \rangle$, $i_B = \langle m_B, q_B \rangle$, $\kappa_A = \langle \rho_A, \epsilon_A, \psi_A, \eta_A,$ $p_A, tr_A \rangle$, and $\kappa_B =\langle \rho_B, \epsilon_B, \psi_B, \eta_B, p_B, tr_B \rangle$. For the sake of simplicity we assume that the currency type $cType$ and the unit type $uType$ are the same for all services.

\subsubsection{Sequential composition $A \gg B$}
The sequential composition of the \emph{ConfiguredServices} $A$ and $B$ is the tuple $\langle \Lambda_{A \gg B}, \alpha_{A \gg B}, \beta_{A \gg B}, \sigma_{A \gg B}\rangle$, whose components are defined below.
\begin{itemize}
\item \textbf{Parameters}: $\Lambda_{A \gg B}$
\begin{itemize}
\item Input parameters: $\Lambda_{input(A\gg B)} = \Lambda_{input(A)} \cup (\Lambda_{input(B)} \setminus \Lambda_{output(A)})$, defined as the union of the input parameters of $A$, and input parameters of $B$ that are not output parameters of $A$. 
\item Output parameters: $\Lambda_{output(A\gg B)} = \Lambda_{output(A)} \cup \Lambda_{output(B)}$, defined as the union of the output parameters of $A$ and $B$.
\end{itemize}
\item \textbf{Attributes}: $\alpha_{A \gg B} = \alpha_A \cup \alpha_B$
\item \textbf{Context}: For \emph{ConfiguredServices} $A$ the context is $\beta_A = \langle r_A,c_A \rangle$. This means that $r_A$ is true in context $c_A$ in order that $A$ may be provided. Once the service $A$ has been provided, the context and rules that are true in that context should be computed. Letting these rules $r'_A$ and the context $c'_A$, we need to merge them with $r_B$ and $c_B$, $\beta_B = \langle r_B,c_B \rangle$ to arrive at $\beta_{A \gg B}$. With this rationale, we define
$\beta_{A \gg B} = \langle r_{A \gg B}, c_{A \gg B} \rangle$, $r_{A \gg B} = r'_A \cup r_B$, and $c_{A \gg B} = c'_A \sqcup c_B$, the smallest closure of contexts $c'_A$ and $c_B$. It is expected that $c'_A \sqsubset c_B$ holds for most of the applications, because anything outside of $c_B$ can be ignored. The semantics of context union ($\sqcup$) and sub-context ($\sqsubset$) and a detailed discussion of context calculus can be found in~\cite{wan06}.
\item \textbf{Contract}: $\sigma_{A \gg B} = \langle f_{A \gg B}, \kappa_{A \gg B}, l_{A \gg B} \rangle$, where
\begin{enumerate}
\item \textbf{function}: $f_{A \gg B} = \langle g_{A \gg B}, i_{A \gg B}, pr_{A \gg B}, po_{A \gg B} \rangle$, $g_{A \gg B} = \langle n_{A \gg B}, d_{A \gg B}, u_{A \gg B} \rangle$, $i_{A \gg B} = \langle m_{A \gg B},$ $q_{A \gg B} \rangle$, where

\begin{tabular}{lcll}
$g_{A \gg B}:$ & & & \\
$n_{A \gg B}$ &=& $n_A \frown n_B$ & naming convention\\
$d_{A \gg B}$ &=& $d_A \cup d_B$  & combine input data parameters\\
$u_{A \gg B}$ &=& $\{u_A,u_B\}$ & both function addresses are necessary\\\\
$i_{A \gg B}:$ & & &\\
$m_{A \gg B}$ &=& $m_A \frown m_B$ & naming convention\\
$q_{A \gg B}$ &=& $q_A \cup q_B$ & combine output parameters \\\\
$pr_{A \gg B}$ & =& $pr_A \cup (pr_B \setminus po_A)$ & if $B$ requires more constraints\\
$pr_{A \gg B}$ & =& $pr_A$ & if $B$ does not require more constraints\\
$po_{A \gg B}$ &=& $po_A \cup po_B$ & if $po_A$ is observable\\
$po_{A \gg B}$ &=& $po_B$ & if $po_A$ is not observable\\
\end{tabular}
\item \textbf{Nonfunctional Properties}: $\kappa_{A \gg B}= \langle \rho_{A \gg B}, \epsilon_{A \gg B}, \psi_{A \gg B}, \eta_{A \gg B}, p_{A \gg B}, tr_{A \gg B} \rangle$ where,
\begin{itemize}
\item Safety (timeliness): $\rho_{A\gg B} = \rho_{A} + \rho_{B}$.
\item Safety (data): $\rho_{A\gg B} = \rho_{A} \cup \rho_{B}$.
\item Security: $\epsilon_{A\gg B} = \epsilon_A \cup \epsilon_B$.
\item Availability: $\eta_{A\gg B} = \eta_A + \eta_B$.
\item Reliability: $\psi_{A\gg B} = Min(\psi_A,\psi_B)$.
\item Price: $p_{A\gg B} = \langle a_{A\gg B}, cu_{A\gg B}, un_{A\gg B}\rangle$ where $cu_{A \gg B}$ = $cu_{A}$ = $cu_{B}$, $un_{A \gg B}$ = $un_{A}$ = $un_{B}$, and  
$$
a_{A \gg B} = \left\{ \begin{array}{rl}
  a_A + a_B &\mbox{normal pricing} \\
  max\{a_A,a_B\} &\mbox{promotional}\\
  min\{a_A,a_B\} &\mbox{special sale}
       \end{array} \right.
$$
\item Provider Trust: Let $tr_{A\gg B} = \langle ce_{A\gg B}, pg_{A\gg B}, re_{A\gg B}\rangle$. Given a set $s_t$ of trust values, it should be possible to define $avg (s_t)$, $choose(s_t)$, $glb(s_t)$, and $lub(s_t)$ which respectively computes the average, selects randomly one value, and computes the least and greatest values from the set $s_t$. Any one of these functions may be used by the Service Provider in providing $ce$ and $re$. Each choice has some significance. Choosing $avg$ reflects `unbiased views of customers', choosing $choose$ reflects a randomly selected customer opinion, choosing $glb$ reflects a conservative estimate, and choosing $lub$ reflects the optimistic opinion of customers. For illustration, we use the function $glb$. We compute the trust sets as

\begin{tabular}{lcll}
$ce_{A \setminus B}$ &=& $\{(a,b) \mid (a,b) \in ce_A, (a,b) \notin ce_B \}$ & recommendation \\
&&&given for $A$ only \\
$ce_{B \setminus A}$ &=& $\{(a,b) \mid (a,b) \notin ce_A, (a,b) \in ce_B \}$ & recommendation \\
&&&given for $B$ only \\
$ce_{A \cap B}$ &=& $\{(a,b) \mid (a,b_1) \in ce_A, (a,b_2) \in ce_B, b = glb(b_1,b_2) \}$ & recommendation \\
&&&given for $A$ and $B$\\
\end{tabular}
%\noindent

Similar sets for $re$ are defined. The trust for the composition $A \gg B$ can be defined for different semantics. 
\begin{itemize}
\item {\em Business Logic: Service $A$ is required for service $B$}. In this situation the expectation is that those who bought service $B$ should have obtained service $A$, and hence they bought the service $A \gg B$. That is, the recommendation for $B$ dominates. With this semantics we define 

\begin{tabular}{lcl}
$ce_{A\gg B}$ &=& $ce_{A \cap B} \cup ce_{B \setminus A}$ \\
$re_{A\gg B}$ &=& $re_{A \cap B} \cup re_{B \setminus A}$ \\
\end{tabular}
\item {\em Business Logic: Those who bought service $A$ are most likely to buy service $B$}. In this situation buying $A$ is a certainty. Not everyone who bought $A$ may buy $B$. That is, service recommendation for $A$ dominates.  With this semantics we define 

\begin{tabular}{lcl}
$ce_{A\gg B}$ &=& $ce_{A \cap B} \cup ce_{A \setminus B}$ \\
$re_{A\gg B}$ &=& $re_{A \cap B} \cup re_{A \setminus B}$ \\
\end{tabular}
\item {\em Business Logic: Both services are packaged together}: With this semantics the Service Provider has to collect the sets $ce$ and $re$ from clients and organizations for the new service.  
\end{itemize}
%\noindent 
In all above situations

\begin{tabular}{lcl}
$pg_{A\gg B}$ &=& $pg_A \wedge pg_B$\\
\end{tabular}
\end{itemize}
 
\item \textbf{Legal Issues}: $l_{A\gg B} = l_A \cup l_B$, defined as the union of the issues of $A$ and $B$.
\end{enumerate}
\end{itemize}

\begin{example}\label{ex:third}
The sequential composition rule is applied to compute $rs \gg tt$, where the \emph{ConfiguredServices} $rs$ (repair shop) and $tt$ (tow truck) are defined in Example~\ref{ex:zero}. The formal notation of composite \emph{ConfiguredService} is $s_{rs\gg tt}$ = $\langle \Lambda_{rs\gg tt}, \alpha_{rs\gg tt}, \beta_{rs\gg tt}, \sigma_{rs\gg tt}\rangle$, where the tuple components are
\begin{itemize}
\item The CofiguredService parameters set is $\Lambda_{rs\gg tt} = \{(CarBroken,bool),(deposit,double),(CarType,$ $string),(failureType,string),(RequestTruck,bool),(HasAppointment,bool),(numberOfHours,$ $int),(RequestConfi,bool)\}$. 
\item The ConfiguredService attribute set is $\alpha_{rs\gg tt}= \Phi$.
\item The ConfiguredService context is $\beta_{rs\gg tt}$ = $\langle r_{rs\gg tt}, c_{rs\gg tt}\rangle$, where the context rules are $r_{rs\gg tt} = \{(membership==caa)\}$ and the context information is $c_{rs\gg tt} = \{(Location,(Montreal,Canada))\}$.
\item The ConfiguredService contract is $\sigma_{rs\gg tt}$ = $\langle f_{rs\gg tt}, \kappa_{rs\gg tt}, l_{rs\gg tt} \rangle$
\item The contract function is $f_{rs\gg tt} = \langle g_{rs\gg tt}, i_{rs\gg tt}, pr_{rs\gg tt}, po_{rs\gg tt}\rangle$
\item The function signature is $g_{rs\gg tt} = \langle n_{rs\gg tt}, d_{rs\gg tt}, u_{rs\gg tt} \rangle$, where the name is $n_{rs\gg tt}=(ReserveRS\&TT)$, the address is $u_{rs\gg tt} =(XXXYYY)$ and the input parameters are $d_{rs\gg tt}=\{(CarBroken,bool),(deposit,$ $double),(CarType,string),(failureType,string),(RequestTruck,bool)\}$.
\item The function result is $i_{rs\gg tt} = \langle m_{rs\gg tt}, q_{rs\gg tt}\rangle$ , where the result name is $m_{rs\gg tt} = (ResultRS\&TT)$ and the output parameters are $q_{rs\gg tt} =\{(HasAppointment,$ $bool) ,(numberOfHours,int) ,(Request$ $Confi,bool)\}$.
\item The precondition is $pr_{rs\gg tt} = \{(CarBroken==true),(RequestTruck==true)\}$ and the postcondition is $po_{rs\gg tt}=\{(HasAppointment==true),(RequestConfi==true)\}$.
\item The contract legal issues are $l_{rs\gg tt} = \{(deposit=300),(CarType==toyota)\}$.
\item The contract nonfunctional properties are $\kappa_{rs\gg tt} = \langle p_{rs\gg tt} \rangle$, where the price is $p_{rs\gg tt} = \langle a_{rs\gg tt},$ $cu_{rs\gg tt} ,un_{rs\gg tt}\rangle$, the price amount is $a_{rs\gg tt} = ((60*numberOfHours)+100)$, the price currency is $cu_{rs\gg tt} = (dollar)$ and the price unit is $un_{rs\gg tt} = (oneTime)$.
\end{itemize}
\end{example}

\section{Formal Verification}\label{sec:verify}
A service composition consists of multiple interacting \emph{ConfiguredServices} that provide a functionality to meet a specific set of requirements. It is essential to verify that the functional behavior of the service composition meets the requirements of the service requesters while taking into consideration the nonfunctional, legal and contextual conditions. Instead of defining a new verification tool to verify the service composition we follow a transformation approach. In this approach, a formally defined service composition can be automatically transformed into a model understood by an available verification tool that can then be used to perform the formal verification. The goal in our research is to use different verification tools in order to verify a wide range of properties and target different kinds of systems. This is because different verification tools differ in their requirements and abilities. In this paper, we define the transformation rules to generate a model that can be verified using UPPAAL~\cite{tut04} model checking tool.

A full account of UPPAAL language and tool can be found in~\cite{tut04}. In essence, UPPAAL extends the definition of {\em timed automata} (TA) with additional features. The features that are relevant to this paper are (1) \textbf{Templates} that represent TAs with optional parameters and local variables; (2) \textbf{Global variables and user defined functions}, that are introduced in a global declaration section, and shared by all templates; (3) \textbf{Binary synchronization} that forces two TAs to have a synchronized transition caused by an event; (4) \textbf{Channel} that models an input event (labeled with ?) or an output event (labeled with !) in a synchronous transition; (5) \textbf{Committed Location} that models a state where time is not allowed to pass, and allowed to have an outgoing edge; (6) \textbf{Expressions} that include \emph{Guard expressions} involving variables and clock variables to restrict transitions, \emph{Assignment expressions}, which are used to set values of clocks and variables, and \emph{Invariant expressions}, which are defined at locations to specify conditions that should be always true; and (7) \textbf{Edges} denoting transitions between locations. An edge specification consists of the four expressions 1) \emph{Select}, which assigns a value from a given range to a defined variable, 2) \emph{Guard}, an edge is enabled for a location if and only if the guard is evaluated to true, 3) \emph{Synchronization}, which specifies the synchronization channel and its direction for an edge, and 4) \emph{Update}, an assignment statement that resets variables and clocks to required values. UPPAAL can check {\em safety}, {\em reachability}, and {\em liveness} properties that are expressed in TCTL~\cite{HT04}.

\subsection{Transforming the Service Composition into UPPAAL TA}\label{sec:rules}
This section presents the rules for transforming a service composition into a UPPAAL TA. Let $S=\{s_1,...,s_n\}$ be the set of \emph{ConfiguredServices} to be composed. Let $\Upsilon$ be the execution flow defining the composition, and $SC =\langle S, \Upsilon, \Lambda, \alpha, \beta, \sigma \rangle $ be the resulting composition. Let $TA = \langle L,L_0,K,A,E,I\rangle $ be the definition of a UPPAAL TA, where $L$ is a set of \emph{locations} denoting the states, $L_0$ is the \emph{initial} state, $K$ is a set of \emph{clocks}, $A$ is a set of \emph{actions} that cause transitions between locations, $E$ is a set of \emph{edges}, and $I$ is a set of \emph{invariants}. The transformation rules will construct $T = \{ta_1,...,ta_n\}$, a set of UPPAAL templates. The first step is to define the following in the global declaration section in UPPAAL.
\begin{enumerate}
\item Two channel variables are defined for each $s_i$. The first represents the request and the second represents the response.
\item A Boolean variable is defined for every precondition and input parameter in $SC$ and assigned to \emph{true}. These variables are used to verify if preconditions and input parameters exist before execution.
\item A Boolean variable is defined for every postcondition and output parameter in $SC$ and assigned to \emph{false}. These variables are used to verify if postconditions and output parameters exist after execution.
\item A typed variable is defined for every parameter in $SC$. The type can be any simple type, such as {\tt int}, or a structured data type.
\item The following variables of type {\tt double} are defined and assigned to 0 for each composition flow:
\begin{itemize}
\item \emph{PathPrice}, which represents the total price of the composition flow.
\item \emph{PathAvailability}, which represents the availability of the composition flow.
\item \emph{PathReliability}, which represents the reliability of the composition flow.
\item \emph{PathTime}, which represents the safety time guarantee of the composition flow.
\end{itemize}
\item Boolean variables representing the elements of the legal issues are defined. These variables are used in defining the Legal issues as Boolean statements.
\item A UPPAAL structure that represents the contextual information of the service requester is defined. The structure contains dimensions and associated tag values.
\end{enumerate}

\subsubsection{Transformation Rules}
The transformation rules are divided into two sets. The first set defines the rules to transform an individual \emph{ConfiguredService} into a TA. The second set defines the rules to transform the composition flow into a TA. Each \emph{ConfiguredService} can be mapped to a UPPAAL template in a one to one manner. A \emph{ConfiguredService} $s_i$ = $\langle \Lambda_i, \alpha_i, \beta_i, \sigma_i\rangle $ is mapped to a template $ta_i= \langle L_i,L_{0i},K_i,A_i,E_i,I_i\rangle $. Following are the transformation rules to generate $ta_i$ for each $s_i$.
\begin{enumerate}
\item For each $ta_i$ create two locations $L_i = \{l_1, l_2\}$, and set the first location as the initial state $L_{0i} = \{l_1\}$.
\item Create two edges in $E_i = \{e_1, e_2\}$ in $ta_i$, with edge $e_1$ directed from $l_1$ to $l_2$ and edge $e_2$ directed from $l_2$ to $l_1$.
\item Define an action for each $s_i$ and add it to $A_i$.
\item Add to edge $e_1$ the following expressions:
\begin{enumerate}
\item Add to guard the condition that all $s_i$ preconditions are equal to true.
\item Add to guard the condition that all $s_i$ input parameters are available.
\item Add to guard the condition that the $s_i$ contextual rules are satisfied.
\item Add to guard the condition that the $s_i$ legal rules are satisfied.
\item Add to Sync the channel variable corresponding to $s_i$ request and follow it with ?.
\end{enumerate}
\item Add to edge $e_2$ the following expressions:
\begin{enumerate}
\item Add to update the statement that assign all $s_i$ postconditions variables to true.
\item Add to update the statement that assign all $s_i$ output parameters variable to true.
\item Add to Sync the channel variable corresponding to $s_i$ responses and follow it with !.
\end{enumerate}
\end{enumerate}

\noindent The steps described above generates a TA for each \emph{ConfiguredService}. The next step is to generate the main TA that maps to the composition execution flow. Before generating this TA, the composition flow should be flattened to contain only sequential composition construct $\gg$. In essence, every composition flow can be flattened into a set of sequential composition flows of \emph{ConfiguredServices}~\cite{NAM11}. 

\begin{example}\label{ex:fourth}
The composition $(A\triangleright_{c1} B)\gg (C||D)\gg F_{\circ_{c2}}$ defined in Example~\ref{ex:second} can be flattened into 8 composition flows, where $X_c$ indicates that $X$ is associated with condition $c$. These are: (1) $A_{c1}\gg C\grave{\gg} D$, (2) $A_{c1}\gg C\grave{\gg} D\gg F_{c2}...\gg F_{c2}$, (3) $A_{c1}\gg D\grave{\gg} C$, (4) $A_{c1}\gg D\grave{\gg} C\gg F_{c2}...\gg F_{c2}$, (5) $B_{\neg c1}\gg C\grave{\gg} D$, (6) $B_{\neg c1}\gg C\grave{\gg} D\gg F_{c2}...\gg F_{c2}$, (7) $B_{\neg c1}\gg D\grave{\gg} C$, and (8) $B_{\neg c1}\gg D\grave{\gg} C\gg F_{c2}...\gg F_{c2}$.
\end{example}

\noindent The main TA will contain an idle state. For each flattened composition flow, a path of states is created in the main TA starting from this idle state according to the following rules.
\begin{enumerate}
\item For each \emph{ConfiguredService} create two states. The first represents the request for the \emph{ConfiguredService} and the second represents the completion of the execution. 
\item For each \emph{ConfiguredService}, if it contains a safety time constraint, create a new clock and add the timing constraint as an invariant on the location. Exception: if the sequential construct resulted from parallel flattening $X\grave{\gg} Y$, only add the invariant to the state with the highest time constraint of $X$ and $Y$, and make the other state a committed state. 
\item For each \emph{ConfiguredService} create two edges. The first connects the state representing the previous \emph{ConfiguredService} in the flow, except for the first \emph{ConfiguredService} where it connect idle state, to the first state defined in rule 1. The second connects the first state to the second state of rule 1.
\item If the \emph{ConfiguredService} is associated with a condition (conditional choice or iteration condition), add this condition as a guard statement on the first edge of rule 3.
\item If the \emph{ConfiguredService} has a safety data conditions, add this condition as a guard statement on the first edge of rule 3.
\item If the \emph{ConfiguredService} has a price, add to the second edge of rule 3 an update statement that adds the price to the path price variable.
\item If the \emph{ConfiguredService} has an availability nonfunctional property, add to the second edge of rule 3 an update statement that adds the availability to the path availability variable.
\item If the \emph{ConfiguredService} has a reliability nonfunctional property, add to the second edge of rule 3 an update statement that adds the reliability to the path reliability variable. Exception: if the sequential construct resulted from parallel flattening, the update statement is only added to the edge with the highest reliability time.
\end{enumerate}
\noindent A reasoned justification for the transformation steps is given in ~\cite{NAM11}.

\subsection{Verification}
Using UPPAAL editor, the \emph{ConfiguredServices} and their composition are specified as UPPAAL templates following the automatic transformation rules defined in Section~\ref{sec:rules}. UPPAAL verifier can be used to verify the following properties.
\begin{itemize}
\item \textbf{Context}: The context rules are not contradictory, and are met for each \emph{ConfiguredService}.
\item \textbf{Functionality}: The behavior of the composition is correct with respect to functionality, which includes verifying.
\begin{itemize}
\item The preconditions of each participating \emph{ConfiguredService} are met before invocation.
\item The input parameters of each participating \emph{ConfiguredService} are available before invocation.
\item The composition generates the required postconditions and output parameters.
\end{itemize}
\item \textbf{Nonfunctional and trustworthiness properties}: The behavior of the composition is correct with respect to nonfunctional properties, which includes verifying.
\begin{itemize}
\item The composition price is greater than or equal the price of any possible execution flow.
\item The composition safety time constraint is greater than or equal the time constraint of any possible execution flow.
\item The composition availability time is greater than or equal to the availability time of any possible execution flow.
\item The composition reliability time is greater than or equal to the reliability time of any possible execution flow.
\end{itemize}
\item \textbf{Legal issues}: The legal rules are not contradictory, and are met for each \emph{ConfiguredService}. 
\end{itemize}

\begin{example}\label{ex:fifth}
Applying the transformation rules defined above to the service composition $RepairShop\gg TowTruck \gg CarRental$ introduced in Example~\ref{ex:zero}, the composition is transformed into 4 TA's mapped to 4 UPPAAL templates, a template for each ConfiguredService and a template for the composition flow. The TA mapped to the ConfiguredService RepairShop is $ta_{rs}= \langle L_{rs},L_{0rs},K_{rs},A_{rs},E_{rs},I_{rs}\rangle $, as seen in Figure~\ref{fig:rs}, where the tuple components are explained below
\begin{itemize}
\item The set of locations is $L_{rs} =\{ idle, RepairShopProcessing \}$ and the initial location is $L_{0rs} = {idle}$.
\item The set of clocks is $k_{rs}=\Phi$ and the set of invariants is $I_{rs}=\Phi$.
\item The set of actions is $A_{rs} =\{ScheduleApt, AptConfirmed\}$.
\item The set of edges is $E_{rs} =\{(idle-RepairShopProcessing),(RepairShopProcessing-idle)\}$. 
\item The edge connecting 'idle' to 'RepairShopProcessing' has the following statements, where 'parameterB' indicates the variable indicating the availability of the parameter 'parameter': 
\begin{itemize}
\item \emph{Guard}: \verb|(RequesterContext.membership==1)&&(CarBroken==true)&&(car| \verb|Type==toyota)&&carTypeB&&failureTypeB|.
\item  \emph{Synchronous}: \verb|ScheduleApt?|. 
\end{itemize}
The edge connecting \verb|RepairShopProcessing| to \verb|idle| has the following statements: 
\begin{itemize}
\item \emph{Update}: \verb|HasAppoitment=true,NumOfDaysB=true,Deposit=Deposit| \verb|+300|.
\item \emph{Synchronous}: \verb|AptConfirmed!|.
\end{itemize}
\end{itemize}
\begin{figure}[!h]
\begin{center}
	\subfigure[]{\label{fig:rs}\includegraphics[width=0.25\textwidth]{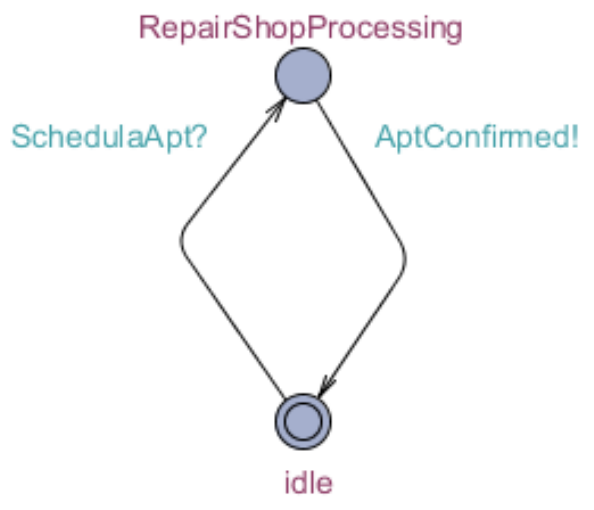}}
	\subfigure[]{\label{fig:tt}\includegraphics[width=0.25\textwidth]{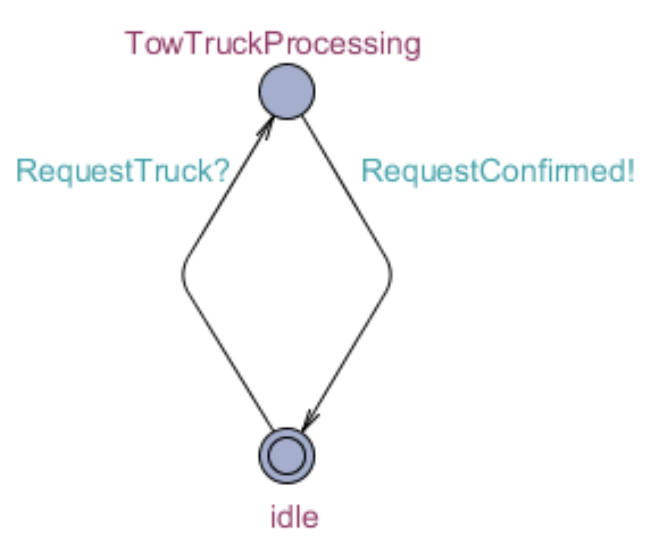}}
	\subfigure[]{\label{fig:cr}\includegraphics[width=0.18\textwidth]{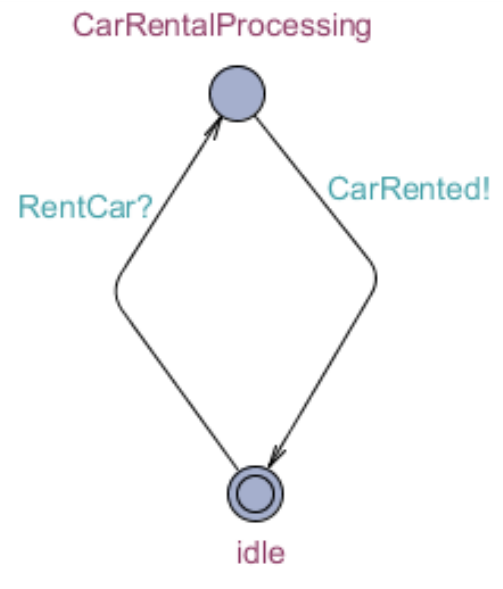}}
\end{center}
  \caption{a) RepairShop TA, b) TowTruck TA, and c) CarRental TA}
  \label{fig:tas}
\end{figure}

The TAs mapped to the \emph{ConfiguredServices} TowTruck and CarRental are created in the same manner. Figure~\ref{fig:example4-main} shows the generated main TA. UPPAAL is used to verify several properties listed below. The notations \verb|M.i| and \verb|M.Final_1| are used to denote the initial and final states of the TA \verb|M|.
\begin{figure}[!h]
	\centering
		\includegraphics[width=0.7\textwidth]
		{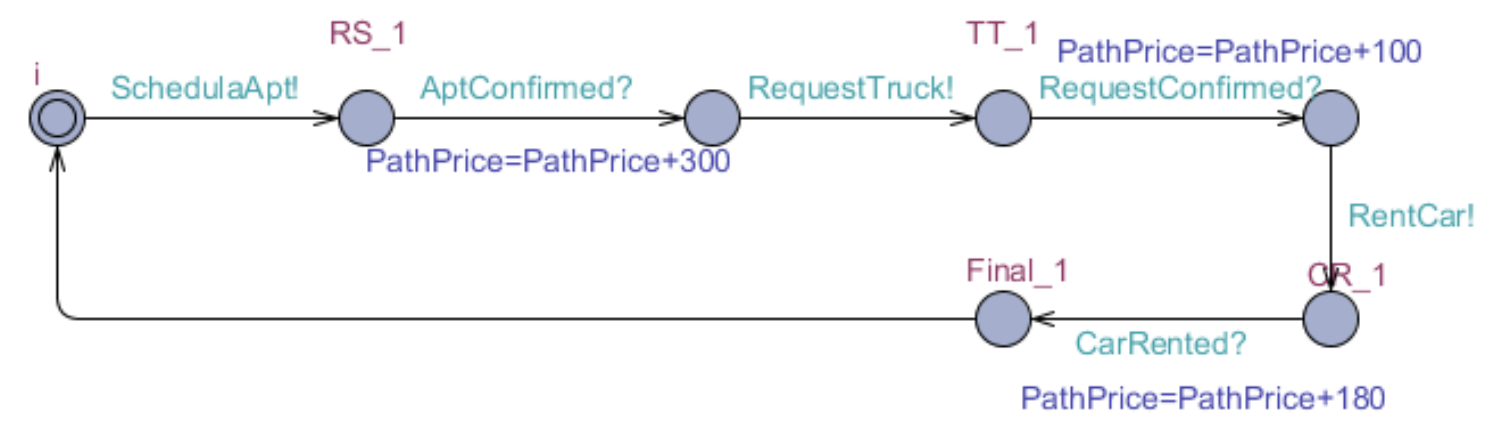}
	\caption{Example~\ref{ex:fifth} Main TA}\label{fig:example4-main}
\end{figure}
\begin{itemize}
\item The composition does not contain any contradiction and can be executed. If the UPPAAL statement \verb|E<> M.Fi| \verb|nal_1| is verified it implies that it is possible to reach the final state of the composition flow. Reaching the final state indicates that all conditions are met and no contradictions exist.
\item The context rules are met. For each context rule an UPPAAL verification condition is generated and verified. For example, \verb|A[] M.i imply RequesterContext.age>=21| is the condition to be verified to assert that the requester is older than 21. Here, \verb|RequesterContext| is the UPPAAL structure holding the contextual information of the service requester. 
\item The composition input parameters are defined before executing the composition flow. For example, \verb|A[] M.i imply failureTypeB| is the condition to be verified in order to assert that the car \verb|failureType| parameter is available before execution. Here, \verb|failureTypeB| is a Boolean variable representing the availability of the parameter \verb|failureType|.
\item The composition output parameters are defined after executing the composition flow. For example, \verb|A[] M.i| \verb| imply !NumOfDaysB| is the condition to be verified in order to assert that the number of days needed to fix the car are not known before executing the composition. The statement \verb|A[] M.i imply !NumOfDaysB|, if verified, asserts that the number of days is known after executing the composition. The parameter \verb|NumOfDaysB| is a Boolean variable representing the availability of the parameter \verb|NumOfDays|.
\item The preconditions are met before executing the composition and the postconditions are met after. For example, \verb|A[] M.i imply| \verb|NeedCar==true| will have to be verified to assert that the precondition ``NeedCar'' is true at the initial state.
\item The composition of nonfunctional properties are correct. For example, \verb|A[] M.Final_1 imply| \verb|firstPathPrice <= 600| will have to be verified to assert that the price of the composite service is less than 600, where 600 is specified as the price of the service composition. 
\item The composition result of the legal rules are correct. For example, \verb|A[] M.Final_1 imply| \verb|400>=Deposit| will have to be verified to assert that the deposit is less than 400, if the legal rule states that ``The service requester should deposit 400 before requesting the service composition''.
\end{itemize}

\end{example}
\section{Related Work}\label{sec:related}
Many researchers, such as~\cite{988756},~\cite{1066866},~\cite{Hinz05transformingbpel}, \cite{1287657} and~\cite{4279619}, have investigated the formal models automata, Petri-net and process algebra as service models and used a transformation approach to arrive at the formal models from service descriptions in one of the languages BPEL~\cite{1214503}, WS-CDL~\cite{ws-cdl} or Orc~\cite{orc}. However, these formal languages can model only the functional behavior of services. Hence, the transformation approaches practiced so far are restricted to only the functionality in composite services, while the nonfunctional, legal and contextual constraints are ignored. As a consequence, the known verification processes cannot be applied to construct composite services in our model. The merit of our work is twofold. One is the introduction of a variety of compositions which can be tailored to the semantics of a business logic, and the other is the ability to combine functional and nonfunctional behavior together with legal and contextual constraints in model checking. 

\section{Conclusion}\label{sec:conc}
Our research aims to define a formal framework for managing and providing service with context-depended contracts. As part of this framework, in this paper we have presented an approach for the formal specification and verification of services with context-dependent contract. We presented a formal definition and a formal composition theory of \emph{ConfiguredServices}. Finally, we presented a formal transformation approach to transform service composition into extended timed automata that can be verified using UPPAAL tool. Currently, we are working on defining a dynamic composition approach that automates the service composition process at execution-time. We are also investigating dynamic reconfiguration issues arising out of defaults and dynamic compositions of services. Finally, we are currently developing a set of tools that automate the composition and verification process.
\bibliographystyle{eptcs}
\bibliography{bibo}

\end{document}